\documentclass[fleqn,usenatbib,useAMS]{mnras}

\usepackage{newtxtext,newtxmath}
\usepackage{natbib} % this is probably into you nips_2018 format file

\usepackage[utf8]{inputenc} % allow utf-8 input
\usepackage[T1]{fontenc}

\usepackage{graphicx}	% Including figure files
\usepackage{amsmath}	% Advanced maths commands
\usepackage{xspace}	
\usepackage{color}
\usepackage{lineno}
\usepackage{tablefootnote}
\usepackage{flushend}
\usepackage{cuted}
\usepackage{natbib}
\usepackage{threeparttable}
\usepackage{xcolor}
\usepackage{float}
\usepackage{dcolumn}
\usepackage{multicol}
\usepackage{adjustbox}
\usepackage{multirow}
\usepackage{caption}
\usepackage{supertabular}
\usepackage{subcaption}
\usepackage{longtable}
\usepackage{caption}

\newcommand{\iitb}{1}
\newcommand{\lsstc}{2}
\newcommand{\iia}{3}
\newcommand{\minnesota}{4}

\newcommand{\caltech}{5}
\newcommand{\MIT}{6}
\newcommand{\texas}{7}
\newcommand{\heid}{8}
\newcommand{\aries}{9}
\newcommand{\nasa}{10}
\newcommand{\maryland}{11}
\newcommand{\hero}{12}

\newcommand{\sw}[1]{\texttt{#1}}
\newcommand{\sxt}{\sw{SExtractor}}
\newcommand{\scorr}{\ensuremath{S_\mathrm{corr}}}
\newcommand{\sqd}{\ensuremath{\mathrm{deg}^2}}
\usepackage{graphicx}	% Including figure files
\usepackage{amsmath}	% Advanced maths commands
\title[GROWTH-India follow up of S190426c]{GROWTH on S190426c II: GROWTH-India Telescope search for an optical counterpart with a custom image reduction and candidate vetting pipeline}
\author[H. Kumar et al.]{Harsh Kumar,$^{\iitb,\lsstc}$\thanks{E-mail: harshkumar@iitb.ac.in}
Varun Bhalerao,$^{\iitb}$
G.C. Anupama,$^{\iia}$
Sudhanshu Barway,$^{\iia}$
Michael W. Coughlin,$^{\minnesota}$
\newauthor
Kishalay De,$^{\caltech, \MIT}$
Kunal Deshmukh,$^{\texas}$
Anirban Dutta,$^{\iia}$
Daniel A Goldstein,$^{\caltech}$
Adeem Jassani,$^{\iitb}$
\newauthor
Simran Joharle,$^{\heid}$
Viraj Karambelker,$^{\caltech}$
Maitreya Khandagale,$^{\iitb}$
Brajesh Kumar,$^{\iia,\aries}$
Divita Saraogi,$^{\iitb}$
\newauthor
Yashvi Sharma,$^{\caltech}$
Vedant Shenoy,$^{\iitb}$
Leo singer,$^{\nasa, \maryland}$
Avinash Singh,$^{\iia, \hero}$
Gaurav Waratkar$^{\iitb}$
\\
$^{\iitb}$Physics Department, Indian Institute of Technology Bombay, Powai, 400 076, India\\
$^{\lsstc}$LSSTC DSFP Fellow-2018\\
$^{\iia}$Indian Institute of Astrophysics, 2nd Block 100 Feet Rd, Koramangala Bangalore, 560 034, India\\
$^{\minnesota}$School of Physics and Astronomy, University of Minnesota, Minneapolis, Minnesota 55455, USA \\
$^{\caltech}$Division of Physics, Mathematics, and Astronomy, California Institute of Technology, Pasadena, CA 91125, USA\\
$^{\MIT}$MIT-Kavli Institute for Astrophysics and Space Research, 77 Massachusetts Ave., Cambridge, MA 02139, USA\\
$^{\texas}$Department of Physics and Astronomy, Texas Tech University, PO Box 41051, Lubbock TX 79409, USA\\
$^{\heid}$Heidelberg University, Grabengasse 1, 691 17, Heidelberg, Germany\\
$^{\aries}$Aryabhatta Research Institute of Observational Sciences, Manora Peak, Nainital - 263 001, India\\
$^{\nasa}$Astrophysics Science Division, NASA Goddard Space Flight Center, Greenbelt, MD, USA\\
$^{\maryland}$Joint Space-Science Institute, University of Maryland, College Park, MD, USA\\
$^{\hero}$Hiroshima Astrophysical Science Center, Hiroshima University, Higashi-Hiroshima, Hiroshima, Japan - 739-8526
}
\date{Accepted XXX. Received YYY; in original form ZZZ}

\pubyear{2022}

\begin{document}
\label{firstpage}
\pagerange{\pageref{firstpage}--\pageref{lastpage}}
\maketitle
\begin{abstract}
S190426c / GW190426\_152155 was the first probable neutron star -- black hole merger candidate detected by the LIGO-Virgo Collaboration.  We undertook a tiled search for optical counterparts of this event using the 0.7m GROWTH-India Telescope. Over a period of two weeks, we obtained multiple observations over a 22.1~$\mathrm{deg}^{2}$ area, with a 17.5\% probability of containing the source location. Initial efforts included obtaining photometry of sources reported by various groups, and a visual search for sources in all galaxies contained in the region. Subsequently, we have developed an image subtraction and candidate vetting pipeline with $\sim 94\%$ efficiency for transient detection. Processing the data with this pipeline, we find several transients, but none that are compatible with kilonova models. We present the details of our observations, working of our pipeline, results from the search, and our interpretations of the non-detections that will work as a pathfinder during the O4 run of LVK.
% We present the follow-up of the `S190426c' event --- detected by LIGO-Virgo Collaboration (LVC) during the first half of third Observing run (O3a). With a poor false alarm rate of 1.4 $\rm {yr}^{-1}$ and a vast 90~\% credible region of 1923.3~$\rm{deg}^2$, this was the first-ever probable NSBH merger event detection by LVC. We present a search for the optical counterpart in form of kilonova (KN) for this event with 0.7~m GROWTH-India Telescope (GIT) performed in close collaboration with Zwicky Transient Facility (ZTF) and the Blanco 4-m telescope equipped with Dark Energy Camera (DECam). GIT covered 17.51 \% probability of localisation by covering 22.11~$\mathrm{deg}^{2}$ area. We highlight our follow-up efforts for S190426c along with the newly developed \sw{ZOGY} based image subtraction and candidate vetting pipeline with $\sim 94\%$ effective efficiency. Under an assumption that kilonova falls in localisation covered by GIT, we put stringent constraints on KNe models.
\end{abstract}


%%%%%%%%%%%%%%%%%%%%%%%%%%%%%%%%%%%%%%%%%%%%%%%%%%

%%%%%%%%%%%%%%%%% BODY OF PAPER %%%%%%%%%%%%%%%%%%

\section{Introduction} \label{sec:intro}
Coalescing compact object binaries are the primary sources of  gravitational waves (GW) for the current ground-based GW detector networks \citep{2015CQGra..32g4001L,2017NCimC..40..120L, 2020arXiv200802921K,Abbott2016, Abbott2017}. Such events have been a subject of great interest in astronomy over the last decade, especially since the first-ever detection of GW by LIGO-Virgo Collaboration (LVC) on September 14, 2015, from a binary black hole merger event~\citep{Abbott2016}. Such merger events are accompanied by electromagnetic emission when at least one of the merger candidates is a suitable mass neutron star~\citep{CUTLER2002, Metzger_2010, Tanaka_2013}. The discovery of the first BNS merger event GW170817 has laid out a robust foundation for these claims \citep{2017Sci...358.1556C, 2017ApJ...848L..12A, 2017Sci...358.1565E, 2017Sci...358.1559K}. LIGO \& Virgo detected this event during Observation run 2 \citep[O2;][]{2017PhRvL.119p1101A}. This event was accompanied by electromagnetic emission spanning the entire spectrum, starting from gamma-ray emission in the form of a short Gamma-Ray Burst GRB~170817A~\citep{2017ApJ...848L..14G, 2017ApJ...848L..13A, 2018MNRAS.478..733L} just $\sim$2~sec after the GW, followed by high and low energy X-rays afterglow emission \citep{2018A&A...613L...1D}. The UV, optical, and IR counterparts in the form of a kilonova~\citep[KN;][]{2017ApJ...848L..24V} were detected hours after the GW signal. At later times, emission was detected at much longer wavelengths in non-thermal radio bands~\citep{article, 2018ApJ...867...18N, 2017Sci...358.1579H, 2017Sci...358.1559K}. This event has proven to be a role model for research in this field over the last few years. To date, this is the only GW event with a confirmed EM counterpart. The near simultaneous detection of GW and short-GRB signals from the GW170817 event ushered in new era of multi-messenger astronomy. The optical and IR observations of the counterpart `AT2017gfo' helped in getting an independent measurement of the expansion rate of the universe \citep{2020NatCo..11.4129C,2019NatAs...3..940H}, 
%Key findings from this particular event involve 
constraining the equation of state~\citep{Radice_2018, 2020Sci...370.1450D}, radius and mass estimation of the neutron stars \citep{Margalit_2017, Rezzolla_2018, 2019MNRAS.489L..91C}, and established that  %This event provides a strong base for the theories predicting 
such merger sites are the factories of the heavy r-process elements in the universe \citep{2017Sci...358.1570D, 2017Natur.551...67P, 2017MNRAS.472..904L}. In order to further understand the physics of such an event, more GW170817--like detections are required in EM bands.

During the first half of the third observing run (O3a), the GW networks detected a gravitational wave event named `S190426c'/GW190426\_152155~\citep[S190426c hereafter;][]{2021arXiv210801045T} with a non-zero probability of the event being a merger of a neutron star and a black hole (NSBH). In search of the optical counterpart of the event, we followed up this event with the GROWTH-India Telescope \citep[GIT;][]{2022arXiv220613535K}, acquiring data for ten nights. We developed our image subtraction and candidate vetting pipeline for the analysis of this data. In this article, we present the follow-up efforts by our team for this particular event and the development of the pipeline. In \S\ref{sec:event}, we discuss the S190426c event and how the source properties were revised over time. Observation strategy of GIT is presented in \S\ref{sec:obs}. \S\ref{sec:processing} highlights our data reduction pipeline, including the newly developed image subtraction and candidate vetting pipeline. In \S\ref{sec:candidates}, we show the candidates discovered --- none of which are consistent with a kilonova. We discuss the implications of these non-detections in the context of various theoretical models. We conclude with a discussion and future outlook in \S\ref{sec:discuss}.
%\todo{ref to sec 5/6}\hk{Done}.

% \rough{`supervised observing mode': mostly robotic, but human oversight for certain tasks}
% \section{Short description of LIGO event S190426c}
% Should be concise, as most of the other papers have this part. So should we avoid rewriting this in our paper?

\section{S190426c}\label{sec:event}
\subsection{Discovery and initial updates}
\begin{figure*}
    \centering
    \subfloat[Initial localisation]{
        \includegraphics[width=0.4\linewidth]{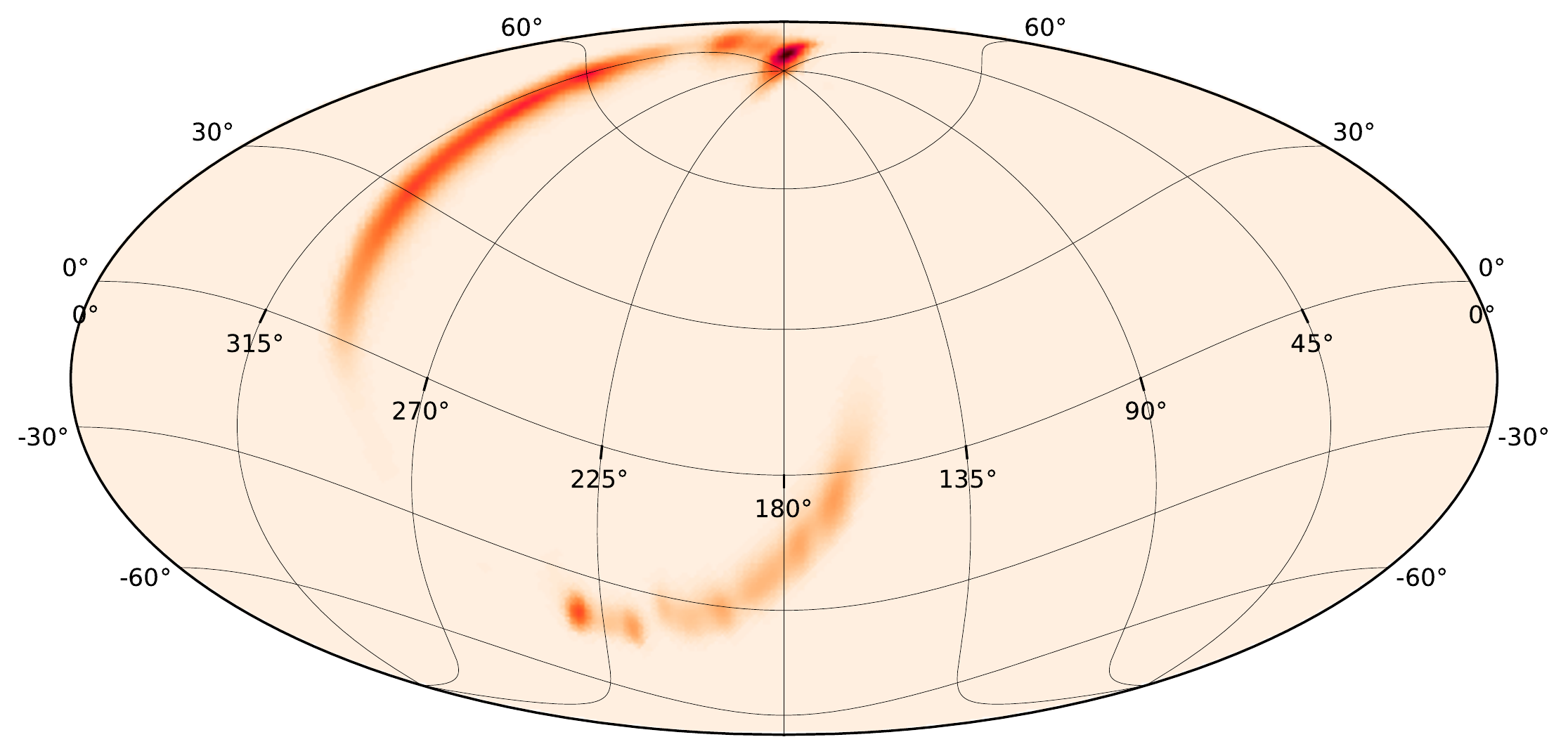} 
        \label{fig:loc_a}
    }
    \subfloat[Updated localisation]{
        \includegraphics[width=0.4\linewidth]{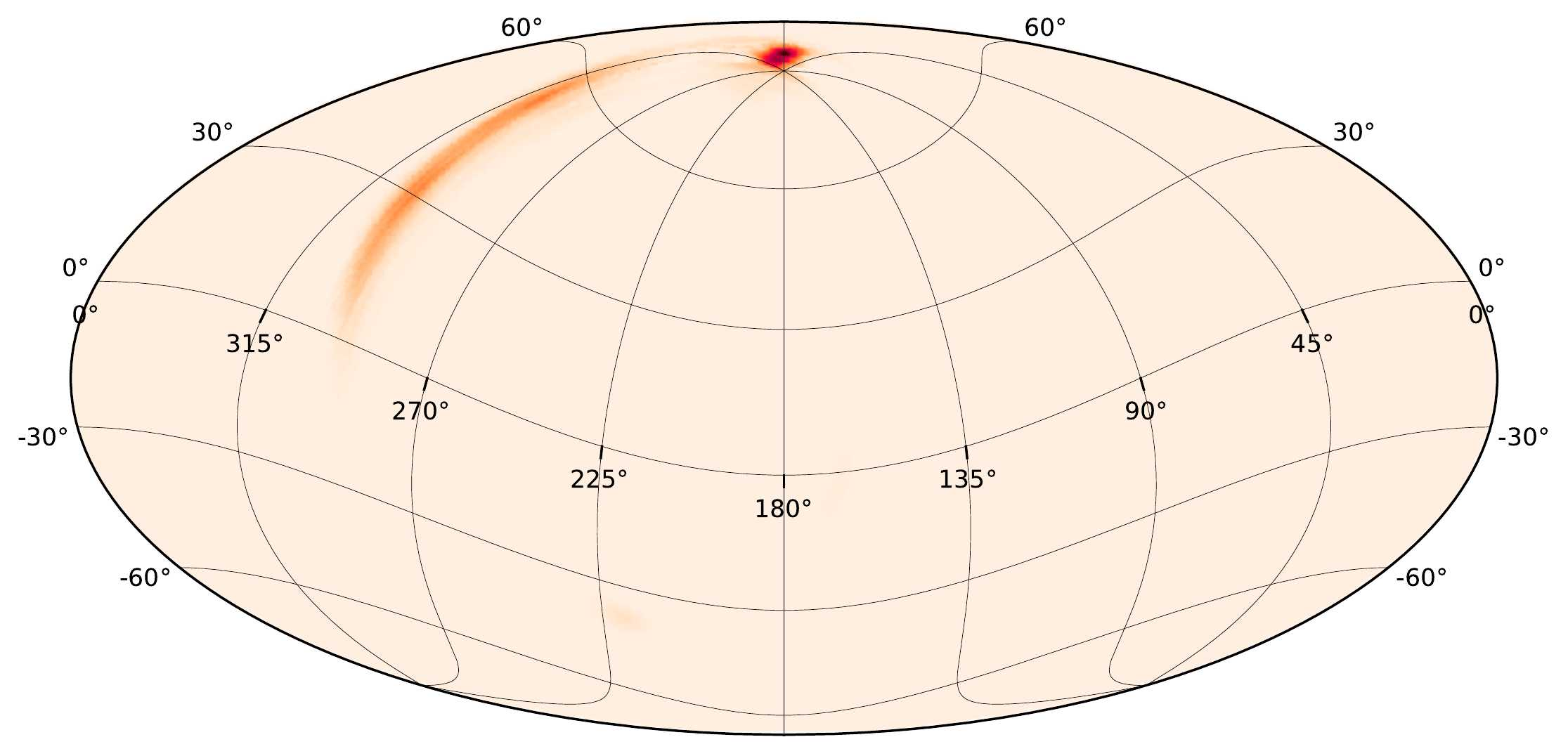} 
        \label{fig:loc_b}
    } \\
    \subfloat[GIT tilling of north pole patch]{
    \centering
        \includegraphics[width=0.80\linewidth]{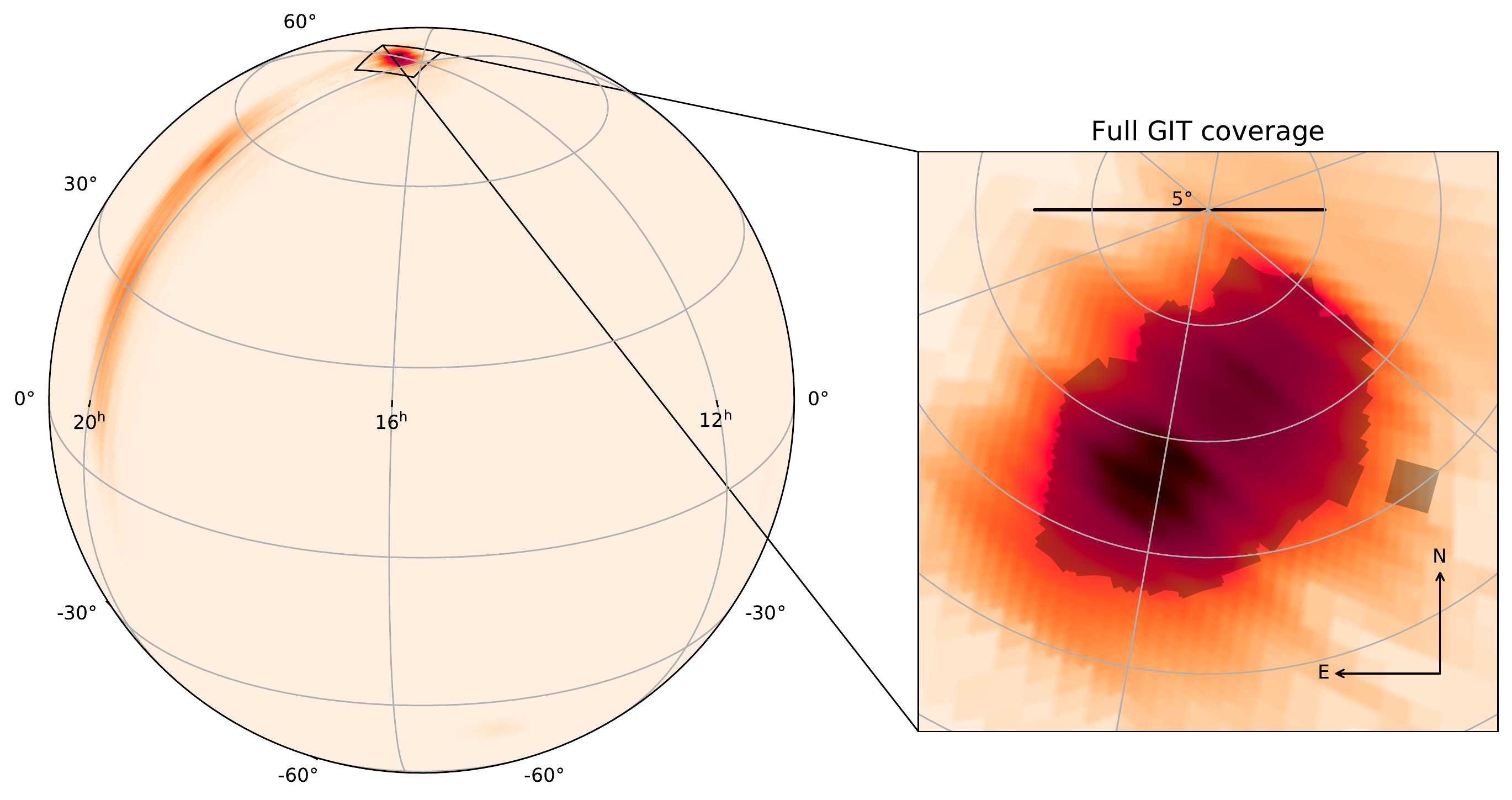} 
        \label{fig:loc_c}
    }\\
    \subfloat[per night tiling of localisation]{
    \centering
        \includegraphics[width=0.85\linewidth]{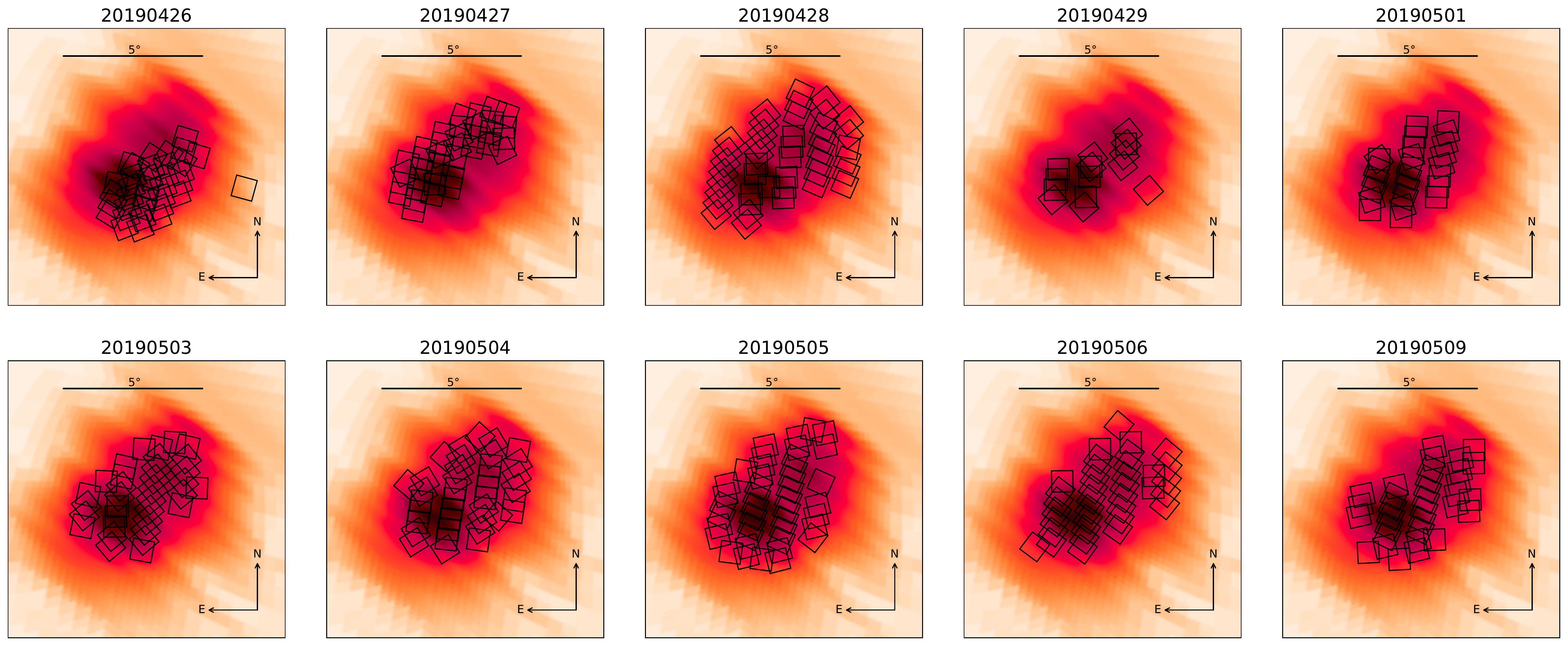} 
        \label{fig:loc_d}
    }
     \caption{ (a) First localisation map circulated by LVC shows that a major portion of localisation probability was in the northern hemisphere with a couple of low probability patches in the southern hemisphere. (b) Updated localisation map by LVC along with GIT tiling indicated by gray shaded region. In updated localisation, the probability gets shifted into the northern hemisphere. (c) S190426c LALInference localisation skymap with GIT tiling shown by squares. (d) Tiles were observed by GIT on each night of observation.  %\todo{\textbf{Replace with the correct figure style}}. \hk{Done in another draft which is prepared for mnras style.}
     }
     \label{fig:localisation}
\end{figure*}

On 2019-04-26 at 15:47:06~UTC, the LIGO Virgo Collaboration issued a VOevent alert~\citep{2006ivoa.spec.1101S} about a binary merger candidate S190426c\footnote{https://gracedb.ligo.org/api/superevents/S190426c/files/S190426c-1-Preliminary.xml,0}. There was a 49\% chance that this was a merger of two neutron stars (Table~\ref{tab:26c_classification}). However, the event had a low statistical significance, with one event per 1.6 years FAR. The source was estimated to be at a distance of 375 $\pm$ 108~Mpc with a 90$\%$ credible sky area of 1262~\sqd\ \citep{2019GCN.24237....1L}. The localisation was divided into three major chunks: a `cap' near the north pole, a long `banana' in the northern hemisphere, and a set of scattered `islands' in the equatorial and southern regions (Figure~\ref{fig:loc_a}). Based on internal discussions within the GROWTH collaboration, it was decided that the GROWTH-India telescope would observe the north polar cap ($\delta \gtrsim 80\degr$), with the Zwicky Transient Facility covering the northern banana and DECam covering the south \citep{2019GCN.24257....1G,2019ApJ...881L...7G}.

The next day, a revised LALInference~\citep{2015PhRvD..91d2003V} sky map was provided, which shrunk the 90\% region slightly to 1131~\sqd, while the luminosity distance estimate remained $377\pm100$~Mpc \citep{2019GCN.24277....1L}. This update removed most of the equatorial and southern localisation regions (Figure~\ref{fig:loc_b}).

\subsection{Nature of the source}
%
% Update on 6 May
Ten days after the event, the event class probabilities were revised \citep{2019GCN.24411....1L}, with a 60\% probability that the source was a Neutron Star -- Black Hole merger (NSBH) and a 15\% chance that it was a binary neutron star (BNS) event. There was a 25\% chance that this was a `MassGap' event, with the class defined such that one of the objects was in the 3--5~$M_\odot$ range.
% In august, they called it terrestrial
Three months after the event, \citet{2019GCN.25549....1L} reported that the event was most likely terrestrial noise with the help of further analysis (Table~\ref{tab:26c_classification}). However, the probability of being astrophysical was non zero.

Final offline analysis of the data \citep{abbott2020gwtc2} shows that the masses of the two components were $\mathrm{m}_1 = 5.7^{+4.0}_{-2.3}$  and $\mathrm{m}_2 = 1.5^{+0.8}_{-0.5}$: a wide span encompassing black holes, neutron stars, and mass gap objects. This is also reflected in the final source class probabilities, which were not explicitly revised in the re-analysis. In our discussion (\S\ref{sec:discuss}), we consider two possibilities for the nature of the source: a BNS merger and an NSBH merger.

%\aks{REWRITE THIS SECTION}
%\outline{S190426c skymap, coverage areas, distance, coordination --- selecting the northern cap}\\
%At 2019-04-26T15:21:55~UTC, the gstlal pipeline of LVC detected the S190426c GW event. This particular event was detected by all three operational detectors, H1, L1 and V1 detectors, with a very poor false alarm (FAR) rate of $1.4 \times 10^{0}~\mathrm{yr}^{-1}$ \citep{abbott2020gwtc2}. Within seconds of detection of the event, LVC sent an automated alert, including a bayestar localisation map. The alert indicated 472.4 and 1932.3~$\rm{deg}^{2}$ sky area with 50~\% and 90~\% credible region \citep{2019GCN.24237....1L}. The initial circular suggested a 49~\% probability that this is a binary neutron star merger candidate, 13~\% chances that this could be an NSBH \footnote{NSBH: One of the companions is a black hole, and the other is a neutron star.} type merger event and 24~\% chances to be a MassGap \footnote{MassGap : One of the object mass is in 3-5  $\mathrm{M_{\odot}}$ range} event. Soon, LVC sends out an updated localisation with a refined area of 262 and 1262~$\rm{deg}^{2}$ for 50~\% and 90~\% credible regions and reduces the distance from 422.71 $\pm$ 127.81~Mpc to 377 $\pm$ 100~Mpc \citep{2019GCN.24411....1L}. The detailed initial and updated classification probabilities are summarised in Table \ref{tab:26c_classification}.

\begin{table}
\caption{Initial and revised classification of S190426c candidate event by LVC. \citep{2019GCN.24237....1L, 2019GCN.24411....1L}. Note that the event is contained in the final catalog, }
\begin{tabular}{ccc}
\hline
\multicolumn{1}{c}{Type} & \multicolumn{2}{c}{Classification probability} \\
\multicolumn{1}{l} {} & Initial & Revised \\
\hline
BNS & $49\%$ & $ 24\%$ \\
NSBH & $13\%$ & $6\% $ \\
MassGap & $24\%$ & $12\% $ \\
Terrestrial & $14\%$ & $58\% $ \\
BBH & $0\%$ & $<~1\% $ \\
\hline
\end{tabular}
\label{tab:26c_classification}
\end{table}%

% BNS=0.493199705189
% NSBH=0.129311795338
% BBH=0.0
% MassGap=0.237362878651
% Terrestrial=0.140125620822
%
%\begin{figure*}
%    \centering
%    \begin{subfigure}[t]{0.45\textwidth}
%        \centering
%        \includegraphics[width=\linewidth]{mollweide_bay.pdf} 
%        \caption{Initial localisation} \label{fig:loc_a}
%    \end{subfigure}
%    \begin{subfigure}[t]{0.45\textwidth}
%        \centering
%        \includegraphics[width=\linewidth]{mollweide_lal.pdf} 
%        \caption{Updated localisation} \label{fig:loc_b}
%    \end{subfigure}
%    \begin{subfigure}[t]{0.90\textwidth}
%    \centering
%        \includegraphics[width=\linewidth]{all_coverage_lal.pdf} 
%        \caption{GIT tilling of north pole patch} \label{fig:loc_c}
%    \end{subfigure}
%    \begin{subfigure}[t]{0.90\textwidth}
%    \centering
%        \includegraphics[width=\linewidth]{grid_5_2.pdf} 
%        \caption{per night tiling of localisation.} \label{fig:loc_d}
%    \end{subfigure}
%     \caption{ (a) First localisation map circulated by LVC shows that major portion of localisation probability was in northern hemisphere with a couple of low probability patches in southern hemisphere. (b) Updated localisation map by LVC. In updated localisation, the probability get shifted into northern hemisphere. (c) S190426c LALInference localisation skymap with GIT tiling shown by squares. (d) Tiles observed by GIT on each night of observation. %The BLAH BLAH colors shows two different tiling schedule for alternate night. \rough{\textbf{Replace with the correct figure.}}
%     }
%     \label{fig:localisation}
%\end{figure*}

%
%The initial skymaps circulated by LVC for the S190426c event consist of three major localisation patches, including a high probability region near the north pole and a long tail in the northern hemisphere with more scattered probability, and a very low probability region in the southern hemisphere. The updated localisation smeared out the majority of the southern patch and shifted the probability more towards the north pole. The initial and final updated localisation map can be found in figure \ref{fig:loc_a} and \ref{fig:loc_b} respectively. \rough{A very recent offline analysis} of S190426c event by LVC estimates component masses of $\mathrm{m}_1 = 5.7^{+4.0}_{-2.3}$  and $\mathrm{m}_2 = 1.5^{+0.8}_{-0.5}$ \citep{abbott2020gwtc2}. 
%This wide range of estimated masses leaves merger type classification wide open with median masses pointing towards a probable NSBH merger.
%
%This range of estimated masses points towards a probable NSBH merger. However, MassGap and BNS merger types cannot be completely ruled out. During the follow-up and analysis process, we took into account both NSBH and BNS mergers types due to the non-zero probability of both these types of merger type classification 
%%lack of strong statements from LVC on the classification type of this event.
%GIT results are compared with detailed NSBH and BNS mergers outcomes in section \ref{sec:discuss}.

\section{Observations}\label{sec:obs}

\subsection{The GROWTH-India Telescope}\label{sec:git}
The GROWTH-India telescope (GIT) is a robotic optical telescope located at the Indian Astrophysical Observatory (IAO) \citep{2002BASI...30..105C, 2008BASI...36..111S} in Hanle, Ladakh\footnote{\url{https://sites.google.com/view/growthindia/}}. It is a 0.7~m planewave CDK700 telescope coupled with a 16 megapixel Andor iKon-XL camera. The telescope design and the wide-format camera together give the telescope a wide field of view (FoV). The high sensitivity non-vignetted area is best approximated as a 0.67~degree diameter circle. The pixel scale is 0.676~arcsec. Our typical limiting magnitude in the $\mathrm{r}^\prime$ band is 20.5 (5-sigma) in 5-minute exposures and 21.0 in 10-min exposures. Since the commissioning of the telescope in June 2018, we have steadily upgraded our software to make it fully autonomous. In early 2019, the telescope was being operated in a semi-autonomous `supervised observing' mode, where remote observers were responsible merely for initiating various batch scripts and intervening only when there were errors. More details on GIT are available at \cite{2022arXiv220613535K}.

\subsection{Observing schedule}\label{sec:schedule}
%\outline{tiling and overlap --- alternate night schedules --- describe alternate-line strategy}

% Relevant GCNs:
%% 25z: https://gcn.gsfc.nasa.gov/other/S190425z.gcn3
% 24191, 24311 - ZTF on 25z
% 24200 - HCT spectroscopy of two 25z candidates
% 24201 - GIT followup of ZTF candidates
% 24304 - GIT photometry of Swift UVOT candidate
%% 26c: https://gcn.gsfc.nasa.gov/other/S190426c.gcn3
% 24258, 24316, GIT tiling
% 24351 - GIT tiling + followup of grawita transient
% 24257, 24268 DECAM
% 24283 ZTF (refers to tiling strategy with GIT)
% 24331 ZTF

When S190426c was first reported, GIT was involved in the follow-up of the previous candidate S190425z \citep{abbott2020gwtc2,2019GCN.24201....1B,2019GCN.24304....1W}. Based on the localisation of the new event, we decided within the GROWTH collaboration that GIT would cover the northern polar cap \citep{2019GCN.24258....1B}, ZTF would cover the northern `banana' \citep{2019GCN.24283....1C,2019GCN.24331....1P} and DECam would cover the southern `islands' \citep{2019GCN.24257....1G,2019GCN.24268....1A,2019ApJ...881L...7G}. Accordingly, we created an observing schedule for GIT from the GROWTH ToO Marshal \citep{2019ApJ...885L..19C, 2019PASP..131d8001C} using the `bayestar.fits' skymap~\citep{2016PhRvD..93b4013S} and obtained 31 $r^{\prime}$ images covering 7.5~\sqd, with 3.9\% probability of containing the GW source as per this original localisation. Using the updated localisation, this probability increased to $\sim6$\%~\citep{2019GCN.24283....1C}.

On subsequent nights, the revised localisation meant that a larger probability region was accessible to GIT for imaging. With the 0.5~\sqd\ field of view of GIT, we carefully planned our observing sequences to maximise science returns. Theoretical models indicate that the optical counterparts to BNS/NSBH mergers will typically evolve on timescales of a couple of days, or longer \citep{Metzger_2010, Roberts_2010, Tanaka_2013, Barnes_2013, 2015MNRAS.450.1777K}. Hence, we divided the north polar cap into two partially overlapping offset grids that would be observed on alternate nights, covering about 10~\sqd\ each \citep{2019GCN.24316....1W,2019GCN.24351....1K}. Observations were scheduled using an implementation of the `Enhanced Array' scheduling algorithm of \citet{Rana_2017}. Over the next two weeks, data were obtained for as many fields of these grids as possible (Figure~\ref{fig:loc_d}). Each point in the showed region was typically observed 4-5 times in our ten observation epochs. Due to the partial overlap in fields within a grid and overlap between the two grids, some parts of the polar cap was observed as many as 10 times during our follow-up. Observations were missed on a few nights: April 30, 2019; May 2,7,8, 2019, due to inclement weather.

%At the time of the initial LVC Gamma-ray Coordinates Network (GCN) notice for S190426c, GIT was observing in the targeted mode for the follow-up of a few candidates from the S190425z event. We rapidly collaborated within GROWTH collaboration and decided to cover up the high probability localisation area around the northern pole, while the other parts of the localisation were covered by ZTF and DECam \citep{2019ApJ...881L...7G}. Soon after the alert, we generated a scheduling plan for GIT through GROWTH Marshal \citep{2019PASP..131c8003K} using the `bayestar.fits' skymap. On the successive day, the localisation was updated with more probability in the northern path. GIT, with a relatively smaller field of view (FOV) compared to ZTF and Decam, was unable to observe the full north cap of localisation in a single night. Therefore, we opted for two different schedules for the alternate night with tile columns alternate among nights.  

\begin{figure}
    \includegraphics[width=\columnwidth]{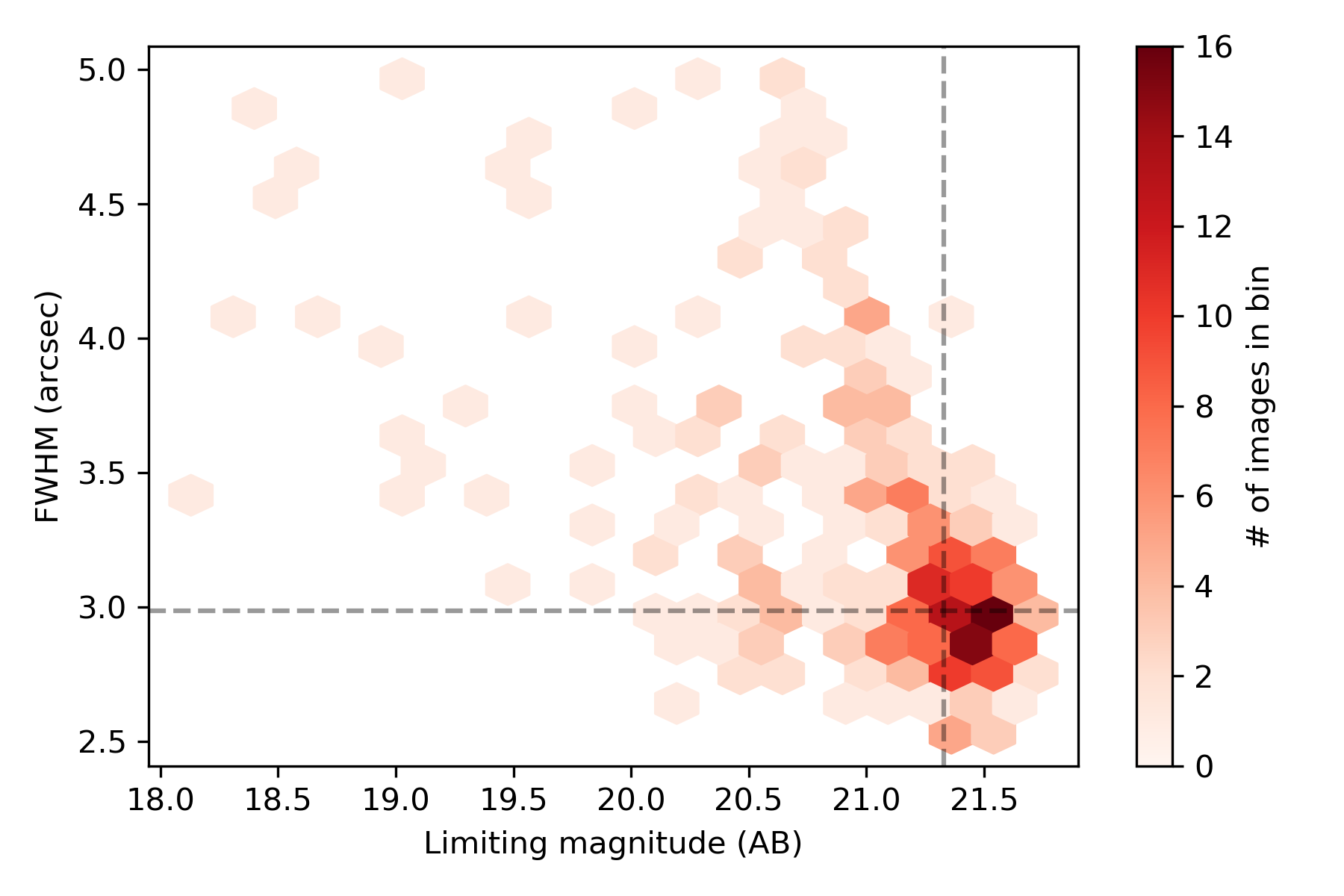}
    \caption{ Statistics of ten nights' data for S190426c. GIT reached a typical depth of 21.32~mag (vertical dotted line) with a median FWHM of 2.98\arcsec during observation (horizontal dotted line).}
    \label{fig:image_stats}
\end{figure}

The primary goal of GIT observations was to identify promising transient candidates, which could then be followed up by the 2-m Himalayan Chandra Telescope \citep[for instance][]{2019GCN.24200....1P} or other GROWTH partners. Hence, we acquired images in a single filter ($r^\prime$) instead of multi filter combination usually preferred in follow-up of such events~\citep{2022ApJS..260...18A}. Given the large median distance of 375~Mpc in the initial LVC alert, we opted to take 600~s exposures, giving us a nightly median limiting magnitude of 20.5 to 21.5 (Figure~\ref{fig:image_stats}) depending on observing conditions. We continued observations for about two weeks to ensure that we would have light curves for any transient candidates and that any event with a late peak would not be lost.

%Total coverage and tiling of alternate nights are shown in \ref{fig:loc_c} and \ref{fig:loc_d}. We scheduled all tiled observations in $r^{\prime}$ filter with a median exposure time of 600~sec. Given the large median distance of 422.71~Mpc to the merger event in the initial alert by LVC,  we opted for an exposure time of ~600sec. The single filter strategy was chosen to prioritise the discovery of possible candidates first and follow them up in all other filters to figure out their association with this particular event. At the time of the event, GIT did not have a robust pipeline to detect transients on the fly. As kilonova typically lasts from a day to a week in the detectable range of current optical facilities \citep{Metzger_2010, Roberts_2010, Tanaka_2013, Barnes_2013, 2015MNRAS.450.1777K}. We continued observation of the same patches for ten nights to avoid losing the vital information of possible kilonova candidate \rough{or interesting transients whose brightness change in a very short timescale.}.

%Algorithm:- Greedy with local tiling 

% We can mention the fast slew of GIT to support the greedy algorithm and also add some spice by saying this one was actually was to go for us.

% Reasons behind the observation strategy and exposure time, the filter is chosen.

\section{Data processing}\label{sec:processing}
Once the Target-of-Opportunity schedule is uploaded to the GIT control computer, it executes the observations and stores data locally at Hanle. The images are compressed using the lossless Rice compression algorithm, using the \sw{fpack} package \citep{pence2011fpack}. They are then automatically downloaded in real-time via satellite link to the CREST campus of the Indian Institute of Astrophysics (IIA), from where another script downloads them to the final processing system at the Indian Institute of Technology Bombay (IITB), where they are uncompressed for further processing.

\subsection{Data reduction}

\begin{figure*}
    \centering
    \includegraphics[width=1.\textwidth]{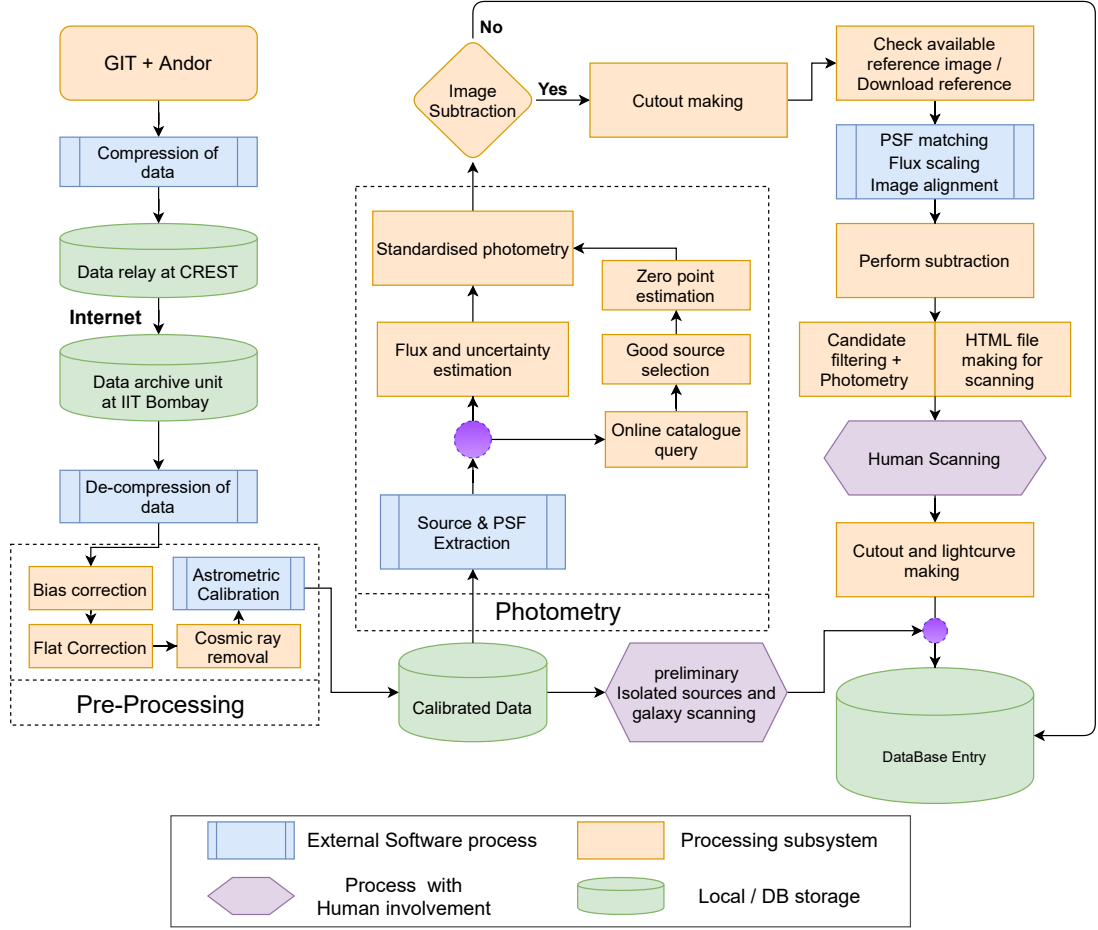}
    \caption{Data reduction pipeline flow of GIT. The light green elements indicate the local or database storage system for data. Light blue boxes are external software dependencies. Light brown colored boxes represent python bases processing subsystems. All processes are described in full details in \S\ref{sec:processing}.
    %\hk{The GIT paper flowchart describe the step for data observation and acquisition.}
    }
    \label{fig:pipe_flow}
\end{figure*}

The GIT data is reduced using the GROWTH-India Image Reduction Pipeline (GRIIPP). The pipeline is divided into three major parts: pre-processing, Point Spread Function (PSF) photometry, and image subtraction. Pre-processing includes generic steps like bias subtraction, flat-field correction, and cosmic-ray removal using standard data reduction techniques. As a last step of pre-processing, astrometry is performed on images using the \sw{solve-field} astrometry engine \citep{Lang_2010}. The corners of the camera extend outside the usable field of the telescope, and we see strong vignetting effects. We limit our analysis to a $\sim 42\arcmin$ square box to exclude regions strongly affected by vignetting. After the pre-processing, images are used for performing PSF photometry as described in \citet{2022arXiv220613535K}. The data reduction steps are depicted in Figure~\ref{fig:pipe_flow}. During reduction of the data obtained for event under discussion in this article, we developed image subtraction and candidate vetting pipeline which has been described in \S\ref{subsec:image_sub}. 
%\hk{The unnecessary details of reduction has been removed. The process of reduction is described briefly so that the flow does not break and flowchart makes more sense.}

% To calculate the zero point of the images, we start by running \sxt\ \citep{1996A&AS..117..393B} on images to extract the sources in the images. The sources which raise \sxt\ flags\footnote{https://sextractor.readthedocs.io/en/latest/Flagging.html} are discarded. 
% We select a subset of `good' sources based on the full width at half maxima (FWHM) of the source and distance from the edge in the GIT images.
% All sources thus obtained are cross-matched with PanSTARRS catalogued sources \citep{chambers2019panstarrs1} via a vizier query \citep{vizier}\footnote{PS1 catalogue accessed through \url{http://cdsarc.u-strasbg.fr/viz-bin/cat/II/349}}. 
% The zero point is calculated from these good sources and appended to the image header.
 
\subsection{Quick-look searches}

GIT had its first light in the Summer of 2018, focusing on automation and reliably acquiring data. As a result, our image subtraction and transient pipelines were not ready when we undertook these follow-up observations. Our real-time processing was limited to two types of `quick-look' searches:

\textit{Search for isolated sources:} We used \sxt\ to find all sources in our images and cross-matched these source lists with publicly available catalogues like PanSTARRS \citep{2016arXiv161205560C} and SDSS \citep{2019ApJS..240...23A} to identify new objects in the images. Only a few significant candidates were found, but they all matched known minor planets in \sw{mpchecker}\footnote{\url{https://minorplanetcenter.net/cgi-bin/checkmp.cgi}} and were rejected for being unrelated to S190426c.

\textit{Search for sources on galaxies:} In case the transient was located on a bright host, it is possible that \sxt\ would not flag it as an independent point source. To cover such cases, we obtained a list of galaxies from the GLADE catalogue \citep{2018MNRAS.479.2374D} and from the NASA/IPAC Extragalactic Database gravitational wave follow-up service\footnote{The service is currently hosted at \url{https://ned.ipac.caltech.edu/uri/NED::GWFoverview/}.}. For instance, 338 GLADE and 8 NED galaxies were present in fields imaged on the first night. We downloaded PanSTARRS thumbnails for each of these galaxies using the \sw{panstamps} utility\footnote{\url{https://github.com/thespacedoctor/panstamps}}, then blinked images in SAOImage DS9 \citep{2003ASPC..295..489J} to look for changes. No transients were found in this search.

\subsection{Image subtraction pipeline}\label{subsec:image_sub}
%\outline{image subtraction}
%\rough{describe the pipeline: first subtraction on galaxies, then on all images}
%\rough{\textit{Transients using image subtraction pipeline}: Later on, image subtraction was performed on all GIT images to find the transients in the field. The image subtraction pipeline has been explained in detail in section \ref{subsec:image_sub}}

We undertake a more rigorous search for transients using the GIT image subtraction pipeline based on the ZOGY algorithm \citep{Zackay_2016}. The pipeline is built using a combination of combination of Astropy modules \citep{2013A&A...558A..33A,2018AJ....156..123A}, \sxt, \sw{PSFEx} \citep{2011ASPC..442..435B}, \sw{SWarp}  \citep{2010ascl.soft10068B}, \sw{SCAMP} \citep{2006ASPC..351..112B}, and \sw{ZOGY} based pipeline \citep{2017zndo...1043973G} to perform subtraction. 

The $\sim0.5$~\sqd\ FoV of GIT makes it infeasible for us to have reference images from our telescope for the entire sky. Instead, we rely on PanSTARRS images  \citep{chambers2019panstarrs1, 2018AAS...23143601F}, downloaded using \sw{panstamps}, as reference images for our processing. 

There are several factors that we need to handle before undertaking image subtraction. PS1 images are limited to a size of about 26\arcmin, and a \sw{panstamps} query returns a cutout that contains the queried coordinate but is not necessarily centred on it. GIT images have an un-vignetted field of 46\arcmin\ and have a different position angle from the PanSTARRS cutouts. Furthermore, ZOGY-based image subtraction is a memory-intensive process, and processing the full 16-megapixel GIT image is infeasible on typical desktop computers. Lastly, there may be non-uniformities in response across the GIT image due to the relatively large image size and FoV. \citet{Zackay_2016} recommend using relatively smaller images to minimise the effects of in-homogeneous transparency and residual astrometric shifts.

As a result, we divide the image into a $4\times4$ grid of cutouts for image subtraction. The cutouts have an overlap of 100~pixels ($\sim 1\arcmin$) to ensure that each source is completely present in at least one cutout. In targeted observing mode, if we are interested in just a particular target in the image, we create a single cutout centred on that target. We then seek a PanSTARRS image for the centre of each cutout. Since we were observing the same part of the sky repeatedly, first, a local query is done to see if the requested images already exist. If not, \sw{panstamps} is used to download the image from the image server.

The next step is to match the GIT cutout to the reference image. We use \sxt\ to extract sources from both science and reference images, using a detection threshold of 5$\sigma$. Then we query the Gaia data release 2 (Gaia DR2) \citep{2018A&A...616A...1G} catalogue for the area covered in the cutout. Using our \sxt\ catalogues and the Gaia DR2 positions of sources, \sw{SCAMP} calculates an astrometric solution and corrects astrometric errors between stars of the two input catalogues and the Gaia catalogue.

Next, we use \sw{SWarp} to subtract the background from the cutout and reference image using a 64-pixel mesh and a filter of three mesh blocks. The images are then resampled to a common plate scale and pixel grid. In this process, fluxes are also re-scaled based on the local ratio of pixel scales, and weight maps are generated. Based on all these, \sw{SWarp} also calculates the variances for images. These are added in quadrature to the Poisson noise estimates from the image to obtain Root-Mean Square images (RMS images) needed as an input for the ZOGY algorithm.

Resampling changes the PSF of the images; hence it needs to be estimated again. We run \sxt\ on these resampled images to create catalogues of bright ($\geq 10\sigma$) sources, which are used by \sw{PSFEx} to create a PSF model. Bright, unsaturated and isolated sources in the cutout and reference images are used to calculate flux scaling and astrometric uncertainties. 

In the final step, the pipeline uses the cutout and reference images, the PSF model, RMS images, and astrometric uncertainties to perform image subtraction and calculate the difference image. Another image called the score-corrected statistics image (\scorr\ image) is generated: local maxima in this image are given the location and statistical significance of the source detected in the difference image. 

The data reduction pipeline is designed so that it can perform actions like stacking of images, subtractions over the full image, and subtraction on the specific targets individually. The pipeline is entirely automatic and takes approximately 2.5~minutes to fully reduce the GIT cutout (with image subtraction) on the current processing unit, which is an Intel(R) Core(TM) i7-6700 CPU running at 3.40~GHz supported by 16~GB of random access memory. The average time is calculated assuming that the PanSTARRS reference images are available in the local database. The pipeline takes an extra 20~sec to download the reference image using \sw{panstamps} if necessary.

% \begin{figure}
%     \centering
%     \includegraphics[width=1.\columnwidth]{candidates_stage_filtering.pdf}
%     \caption{ A graphical depiction of candidates number rejection in each stage of pipeline. The outward arrow shows the candidates thrown out by each cut. \rough{\textbf{Replace with the corrected caption.}}}
%     \label{fig:can_filt}
% \end{figure}

\subsection{Detecting transients in the difference image}\label{subsec:filtering}

\begin{table*}[htbp]
\caption{Filtering process of the good candidates from spurious candidates using vetting cuts. Candidates rejected in each step of vetting cut are listed in third column. Column four represents the candidates survived after each cut.}
\begin{center}
\begin{tabular}{|p{0.45\columnwidth}|p{\columnwidth}|r|r|}
\hline
\textbf{Name} & \textbf{Description} & \textbf{Candidates} & \textbf{Candidates} \\
& & \textbf{rejected} & \textbf{left} \\
\hline
Initial candidate & All sources identified as local maxima in \scorr\ images. & --- & 2,096,938\\
\hline
Faintness cut & Transients that are more than one magnitude fainter than the limiting magnitude of the image are rejected. This cut is designed to be conservative. & 816,816 & 1,280,122\\
\hline
Photometric uncertainty & Candidates with photometric uncertainties $>~1$ mag are rejected. & 125,866 & 1,154,256\\
\hline
Vignetting and edge cuts & We created a binary mask, and rejected sources suffer vignetting (outside a 46\arcmin\ circle) or are too close to cutout edges (5 pixels). Note that cutouts overlap by 100 pixels, so the latter step does not reject any source. & 295,744 & 858,512\\
\hline
FWHM cut & We fit a 1-dimensional Gaussian along the central row and then the central column of each candidate, to measure the FWHM ($F_c$), and compared it with the FWHM of the full image PSF ($F_p$). Only candidates with $0.5 <~F_c / F_p <~ 1.5$ were accepted. & 531,309 & 327,203\\
\hline
XY centre cuts & If the centres of the two Gaussian fits were discrepant by more than 10~pixels from each other, the candidates were rejected. & 73,358 & 253,845\\
\hline
Duplicates and single detections & Since image subtraction was performed on overlapping cutouts, several sources were detected in multiple cutouts from the same parent image. These were merged. Any sources that were detected in only one parent image were rejected. &  246,528 & 7,317\\
\hline
Bright candidate rejection & Given the luminosity distance of S190426c is $\gtrsim 300$~Mpc, it is highly unlikely to have a kilonova candidate to be 16 magnitudes in brightness even with very high ejecta masses \citep{Kasen_2015,2019AA...625A.152B,2020ApJ...897...20Z,2020NatAs.tmp..179A}. & 441 & 6,876\\
\hline
Bright star proximity & Sources within 5\arcsec\ of any stars brighter than 15th  magnitude in PS1 were rejected. & 2,708 & 4,168\\
\hline
Visual inspection & Independent visual inspection by three people. & 4,051 & 117 \\
\hline
Grouping & multiple detections of same object grouped. & --- & 23 \\
\hline
Star / galaxy separation & PS1 star-galaxy check. & 18 & 5\\
\hline
\end{tabular}
\end{center}
\label{tab:cuts}
\end{table*}
% \rough{4168 candidates include multiple detections of same source. Number of unique candidates is smaller.}
%\citep{Kasen_2015, 2019AA...625A.152B, 2020ApJ...897...20Z, 2020NatAs.tmp..179A}. 
%Here is a brief description of these cuts:-
%    \textbf{Magnitude Cut:} We performed PSF photometry on all candidates detection by local maxima finder code. Transients that are one magnitude shallower than the limiting magnitude of the image were rejected \todo{sounds ulta...}. Based on this cut, we rejected 816816 candidates. Candidates with magnitude uncertainty of more than one mag were also rejected.
%    

%    \todo{did we have candidates brighter than 15th mag?}
%    
%    \textbf{Duplicate entries and single detection:} We grouped the entries associated with the same sources, deleted all duplicate entries in the database, and rejected all candidates which have just a single detection in data spanning over ten nights.

% \outline{How the spurious sources were eliminated.}
%
We searched for candidates in the subtracted images and detected the local maxima in the \scorr\ image to detect the transients. Among all local peaks corresponding to transients, we choose transients with corrected score ($\mathrm{S}_{corr}$) $>$ 5. Using these criteria, we found a total of 2,096,938 candidates in all images, with the majority of these found to be artefacts. Cores of very bright stars in the original field show some residuals as they have extra Poisson noise sitting at their centre giving rise to many spurious sources in the difference images. Also, the GIT images suffer from vignetting around the edges contributing to many spurious sources. To eliminate these spurious sources, we developed a filtering process that applies various automated cuts to candidates. The steps used to reject spurious candidates are summarised in Table~\ref{tab:cuts}.

 After automatic cuts, we were left with 4,168 detections scanned manually by three observers independently. A majority of the sources were discarded during manual scanning as those were a result of either bad subtraction or residuals of cosmic rays which did not get removed cleanly. The number of good candidates went down to 23 with 117 detections after the manual scanning. Note that all candidates here had multiple detections --- we would have rejected any objects with just one detection. All candidates had underlying sources associated with them. Therefore, we checked these underlying sources for stellar or non-stellar (galaxy) classification with the help of the PS1 catalogue using the method described in \citet{2014MNRAS.437..748F}. All but five sources were found to be stellar. We performed a standard check on MPC for these five remaining candidates to ensure that none of them is a moving object.

\begin{table}
\centering
    \begin{tabular}{|c|c|c|c|}
    \hline
    \multicolumn{4}{|c|}{\textbf{Efficiency}}\\
    \hline
    \hline
     & Raw & Coverage  & Effective \\
    \hline
    Filter efficacy test & 78.8\% & 84.6\% & 93.1\% \\
    \hline
    Blind test & 82.1\% & 86.7\% & 94.7\% \\
    \hline
    \end{tabular}
    \caption{Detection efficiency of GIT image subtraction and candidate vetting pipeline. Raw efficiency indicates the number of sources retrieved using pipeline with respect to inject sources. Coverage indicate the percentage of image portion coverage available in PS1 for a GIT cutout in a single query using \sw{panstamps}. Note that coverage efficiency increased to $\sim$93.5\% in actual analysis after the tests had been completed. Effective efficiency is defined as the efficiency of GIT pipeline considering 100\% reference image coverage.}
    \label{tab:efficiencytab}
\end{table}

% \begin{figure}[t]
%     \centring
%     \includegraphics[width=0.95\columnwidth]{testing_pipeline_nested_pie.png}
%     \caption{ Detection efficiency of GIT pipeline. The upper panel shows the results from testing on random fields with 3100 injected sources. Sources falling in the common good data portion of GIT and the reference image are termed retrievable. Retrieved sources are the ones which were retrieved back after the subtraction and filtering process. The upper left panel shows raw recovering efficiencies. The Upper right panel shows effective efficiencies (calculated using sources falling in the common good data portion of GIT and reference image). The lower panel shows the blind injection test results for S190426c data: (c) Blind injection raw efficiency. (d) Effective efficiencies calculated using common good data portion.}
%     \label{fig:pipe_testing}
% \end{figure}

% \section{Efficiency of the subtraction pipeline}
% \label{testing_pipe}
% \begin{figure}
%     \includegraphics[width=\columnwidth]{Image_stats_v1.png}
%     \caption{ Statistics of ten nights data for S190426c. GIT reached a typical depth of 21.32~mag (vertical dotted line) with median FWHM of 2.98~arcsec during observation (horizontal dotted line).}
%     \label{fig:image_stats}
% \end{figure}

% \todo{define net efficiency = coverage eff * detection efficiency. Here mainly discuss detection efficiency}

\subsection{Coverage efficiency}\label{sec:coverage_eff}
Some complexity is added to our pipeline due to two factors: 1) the position angle of GIT images is not the same as the reference images, and 2) \sw{panstamps} returns a reference image that contains the queried point (centre of our image), but not necessarily at the centre of the PS1 cutout. As a result, we regularly see `holes' where image subtraction could not be performed as the area was outside the reference image. This problem seems to be exacerbated by the fact that our observations are close to the pole. At the first pass, the holes occupied $\sim 15\%$ of our observed fields, giving us a net `\textit{coverage efficiency}' of $\sim 85\%$. Currently, we have a script that helps us identify such holes, and we re-run the pipeline by downloading reference images for those hole centres, which increase the coverage to $\sim 93.5\%$

\begin{figure*}
    \centering
    \includegraphics[width=1.\textwidth]{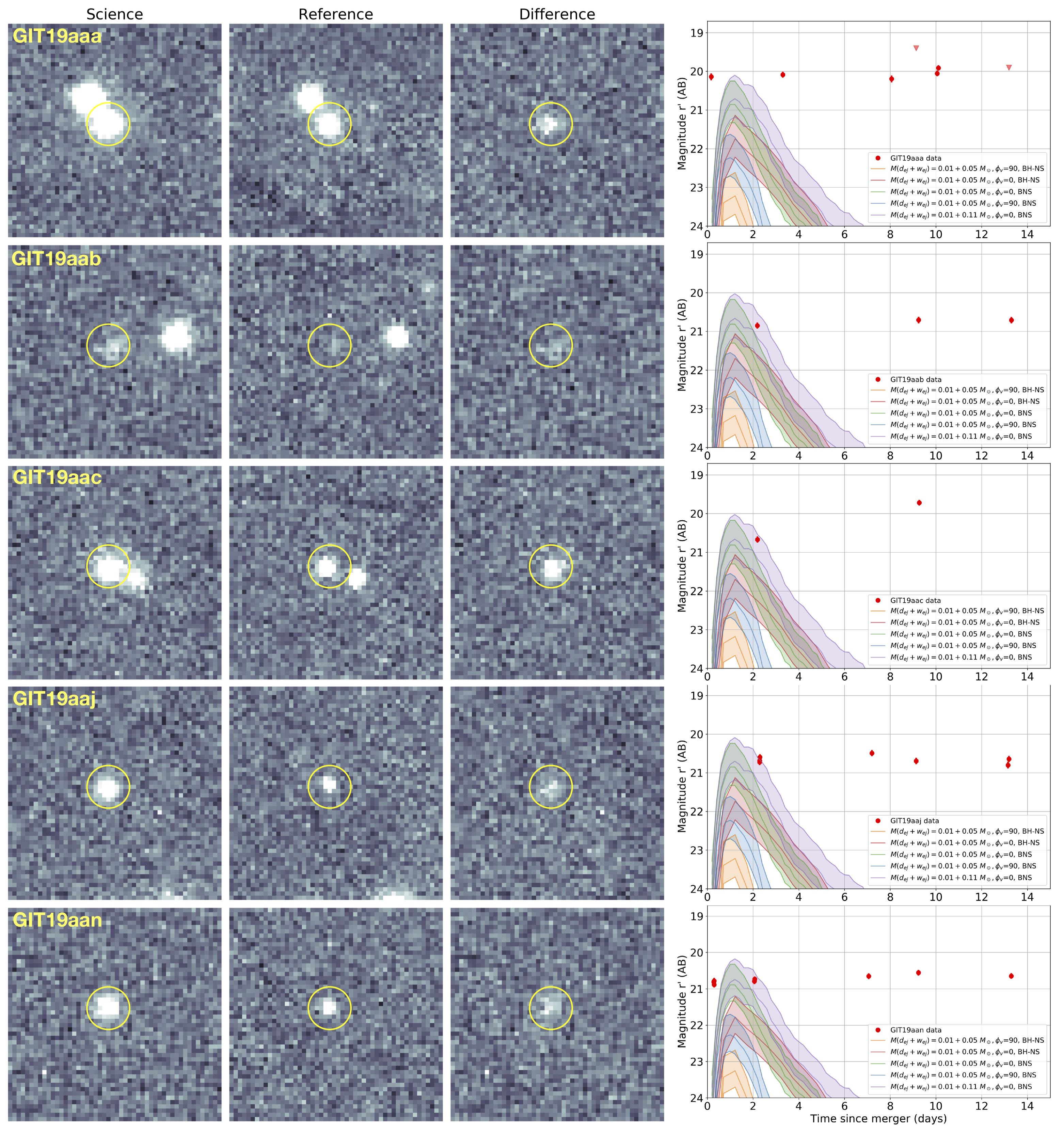}
    \caption{Five good candidates resulted from the GIT data after scanning. The first column represents cutout of candidates from GIT images, second and third columns shows reference from PS1 and difference image respectively. The final columns shows $\rm r^{\prime}$ band light-curve comparison with KN model \citep{2020arXiv201010746H} for each candidate. None of the candidate shows promising light-curve to qualify as kilonova (see \S\ref{sec:candidates}).}
    \label{fig:can_cut}
\end{figure*}

\subsection{Detection efficiency}
 As discussed in \S\ref{subsec:filtering}, the large number of initial candidates were reduced by various filtering steps, followed by human inspection. 
Presently, we lack a machine learning-based real-bogus candidate classifier. 
In order to check the reliability of our procedure, we undertook various tests. 

First, we tested the efficacy of the filtering process before human inspection. We created fake sources using the point spread function of the images, and injected them at random locations in the images. A total of 3,100 sources spanning over magnitude range $\in$[18.5-20.5] were injected across various fields in $\rm g^\prime$ and $\rm r^\prime$ filter images. 476 of these were lost to coverage issues, leaving 2,624 ``retrievable'' candidates. In the actual pipeline, the first candidate identification step post image subtraction included 2,490 sources flagged as candidates in the first step. In addition to the injected sources, about 124,000 spurious sources were flagged as candidates at this stage. After applying the filter criteria, the number of spurious sources decreased drastically to $\sim$4,550, while only 48 injected sources were lost. Thus in the end human scanners would have inspected $\sim7,000$ candidates, and recovered $\sim93\%$ of the injected sources.

Next, we performed a complete end-to-end test ``blind test'' including human scanning, using raw data from our S190426c observations. In each image, we injected between 0--7 sources. Each injected source was repeated in multiple images to test our ability to find multiple detections. The number of repetitions was randomly selected between 4 to 7. The human scanners were unaware of the source locations, magnitudes, and number of repetitions. We obtained comparable results, with an effective efficiency of $\sim 95\%$ (Table~\ref{tab:efficiencytab}). Note that the tests were not repeated after we introduced a script that filled coverage holes (\S\ref{sec:coverage_eff}), which will boost the raw efficiency and hence the raw efficiency by 8--9\%.

We also explored the possibility of using \sxt\ to directly find sources in the difference image. We selected only those candidates which did not raise any \sxt\ FLAGS\footnote{https://sextractor.readthedocs.io/en/latest/Flagging.html} (FLAGS = 0) and with the source FHWM in the range of 0.5 -- 1.5 times the nominal PSF. We found that the detection efficiency for this method is only about 70\%, significantly lower than the ZOGY \scorr\ method.

\section{Results}\label{sec:candidates}
\subsection{Candidates}
Our search process yielded five candidates that passed our filters and had more than one detection each, which we now discuss in detail. First detection images for these 5 candidates along with their full light curves during our observations are shown in Figure~\ref{fig:can_cut}. The figure also shows representative kilonovae lightcurves, which we discuss in \S\ref{sec:KNmodels}
%We compared their light curves to the simulated kilonova model light curves from various theoretical models that are described in details in \S\ref{sec:KNmodels}. 

% For this comparison we used very simusing NSBH and BNS model described in \citet{2020arXiv201010746H} with a most plausible scenario of dynamical ejecta mass $\mathrm{M}(d_ej) \sim0.01~\mathrm{M}_{\odot}$ and post-merger ejecta mass of $\mathrm{M}(d_ej) \sim0.05~\mathrm{M}_{\odot}$ from polar ($\theta_{v} = 90$) and equatorial ($\theta_{v} = 0$) view. We have also looked into the case of very high post-merger ejecta mass of $\mathrm{M}(d_ej) \sim 0.11~\mathrm{M}_{\odot}$ seen from the equatorial ($\theta_{v} = 0$) angle. The KN light curves are scaled for the distance in the direction of the candidate estimated using the \sw{LALInference1.fits.gz,0} skymap circulated by LVC.

 \textit{GIT19aaa}: This candidate was first detected $\sim$ 0.17 days after the event trigger. The candidate was detected four more times on subsequent night observations, with upper limits in two observations. The candidate shows little or no evolution over the two weeks of observations.
% A comparison of KN model light curves to the candidate evolution is shown in the top panel of Figure \ref{fig:can_cut}. The candidate shows very slow (or almost no evolution) throughout 14~days. We rejected this candidate based on slow evolution and comparison with the KNe model light curves.
 
 \textit{GIT19aab}: The field of this candidate was first observed $\sim$ 2.18 days after the event, where we obtained our first detection. The candidate was also detected in subsequent imaging epochs 9.24 and 13.3 days after the trigger, with no non-detections in our full observing period. GIT19aab has nearly a constant magnitude over this timespan.
% The first detection is in agreement with a KN model. However, second and third detection after 9.24 and 13.3 days suggest that the evolution of this candidate is too slow to be a KNe candidate.

 \textit{GIT19aac}: Similar to GIT19aab, this candidate was also observed and detected $\sim$ 2.17 days after the trigger. This candidate brightens by nearly a magnitude over seven days, which is not expected from kilonovae.
% The evolution of this candidate does not show similarities to KNe models implying a disassociation with the S190426c event.
 
\textit{GIT19aaj}: This candidate was first detected $\sim$ 2.28 days after the trigger. It is located in the overlapping area of certain GIT tiles, resulting in multiple observations on some epochs. It has a relatively flat light curve, with signs of intra-night variability. There is an underlying source present at the location of the source, which has a history of variability as per the PS1 Catalogue. Therefore, we conclude that the candidate is likely a result of activity in the underlying source and is not associated with S190426c.

\textit{GIT19aan}: This candidate was detected on the first night of observation itself $\sim$ 0.27 days after the GW event. An almost flat light curve with seven detections over a period of 13 days indicates that this candidate is not associated with the S190426c.

% None of the candidates qualified as a kilonova counterpart for this particular event. Comparing the light curves with KNe models, we can conclude that the ejecta mass $\geq$ 0.12 $\rm M_{\odot}$ is highly unlikely.

In summary, four candidates did not show any significant temporal evolution, while one brightened very slowly. These behaviours are inconsistent with expectations from kilonovae, and we can rule out all of our candidates as potential counterparts to S190426c.

\subsection{Implications of non-detection}\label{sec:KNmodels}

\begin{figure*}
    \centering
    \subfloat[NSBH model for different ejecta mass and viewing angle (absolute).]{
    \centering
        \includegraphics[width=\linewidth]{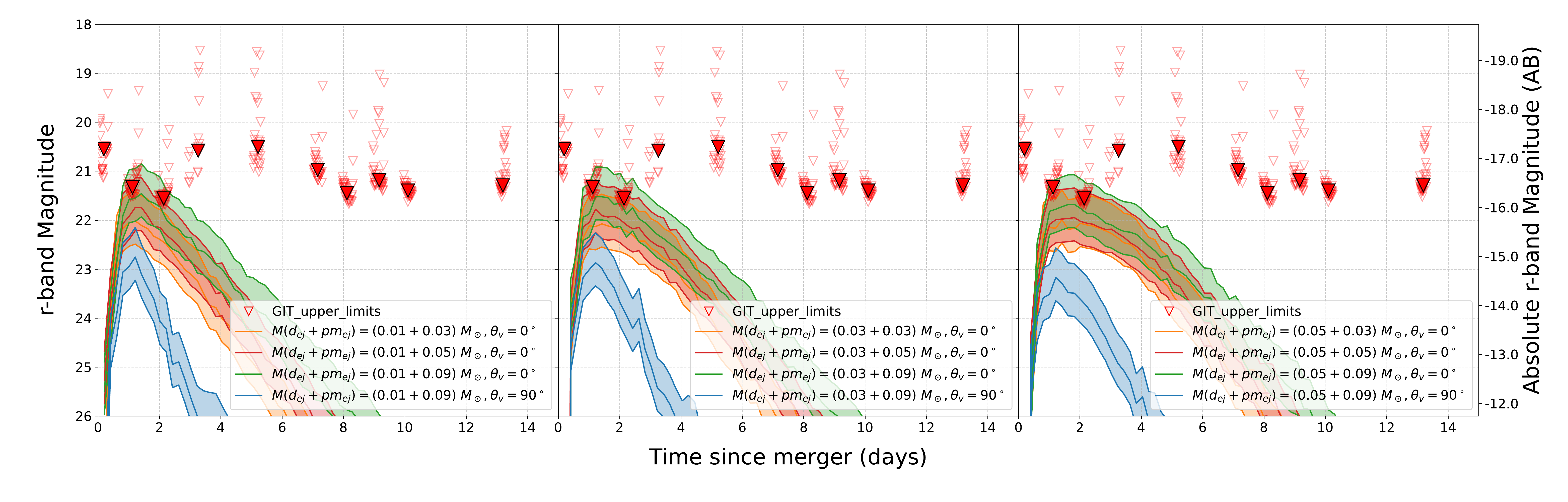} 
        % \caption{
        \label{fig:BH-NS-apr}
    }\\
    \subfloat[BNS model for a few ejecta mass and viewing angle possibilities  (absolute).]{
    \centering
        \includegraphics[width=\linewidth]{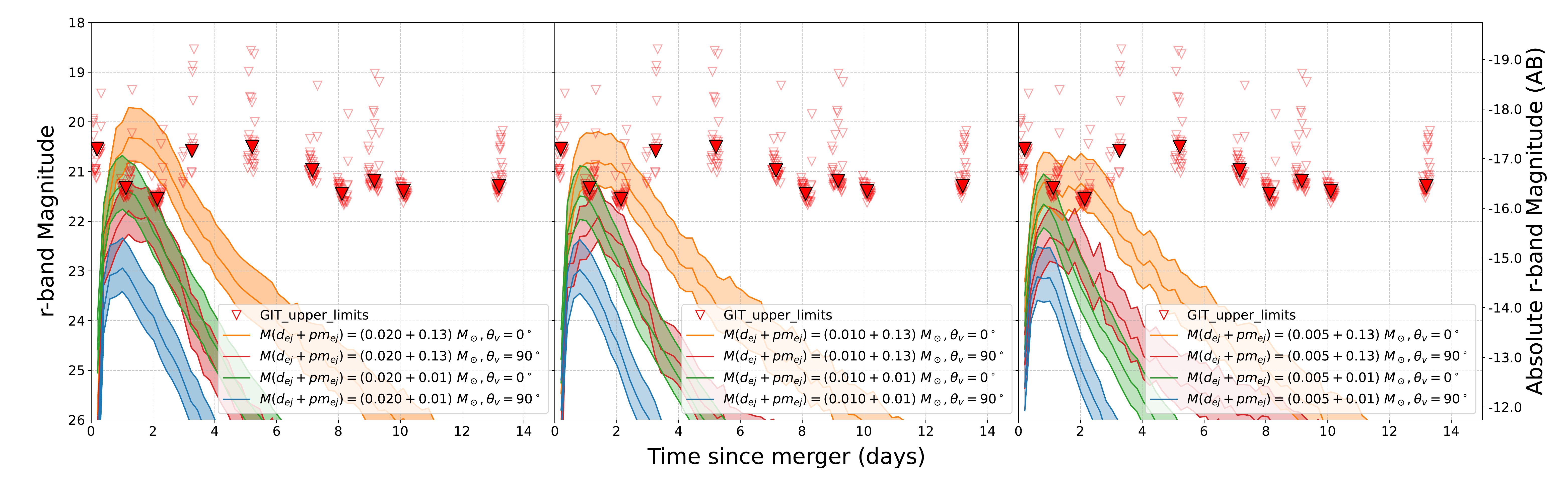} 
        
        % \caption{} 
          \label{fig:BNS-apr}
    }
      \caption{Constraining KN models based on GIT upper limits. The coloured bands indicate the expected range of magnitudes over the one-sigma distance range of the source. The left axis indicates the apparent magnitude, while the right axis shows the source absolute magnitude at the median localisation distance. Different coloured bands show different source models. \textit{Upper panels:} NSBH merger models. \textit{Lower panels:} BNS merger models.
     }
     \label{fig:KNe-models}
\end{figure*}

% \begin{figure*}
%     \centering
%     \begin{subfigure}
%     \centering
%         \includegraphics[width=\linewidth]{BH-NS_model_updated.pdf}
%         \caption{NSBH model for different ejecta mass and viewing angle (absolute).} \label{fig:BH-NS-apr}
%     \end{subfigure}
%     \begin{subfigure}
%     \centering
%         \includegraphics[width=\linewidth]{BNS_model_updated.pdf}
%         \caption{BNS model for a few ejecta mass and viewing angle possibilities  (absolute).}
%       \label{fig:BNS-apr}
%     \end{subfigure}
%     \caption{Constraining KN models based on GIT upper limits. The coloured bands indicate the expected range of magnitudes over the one-sigma distance range of the source. The left axis indicates the apparent magnitude, while the right axis shows the source absolute magnitude at the median localisation distance. Different coloured bands show different source models. \textit{Upper panels:} NSBH merger models. \textit{Lower panels:} BNS merger models.}
% %     \caption{Constraining the NSBH (panel a) and BNS (panel b) models for KNe with the help of GIT upper limits. The inverted triangles denotes the upper limits in $\rm r^\prime$ band scaled to the absolute magnitudes for the median distance of the event. The error bars account for the uncertainty in the distance measurements of the event provided by LVC. The limits are in AB system of magnitudes.}
%      \label{fig:KNe-models}
% \end{figure*}

% 
The initial classification of S190426c indicated that it may be BNS or a NSBH event. Hence, we consider representative theoretical models for both merger types and calculate the expected light curves from both. We compare these to our candidates (Figure~\ref{fig:can_cut}) and also to our non-detection upper limits (Figure\ref{fig:KNe-models}), to constrain the merger ejecta mass. Inspired by \citet{2019MNRAS.489.5037B,dietrich2020new}, we picked a plausible scenario with an ejecta opening angle of 30$^\circ$ for models discussed in this section.

\subsubsection{Neutron Star -- Black Hole merger models}\label{sec:bhnsmodel}
 We compared our $\rm r^\prime$ band upper limits to light curves simulated with \sw{POSSIS}~\citep{2019MNRAS.489.5037B,2020NatAs.tmp..179A} for NSBH models. \sw{POSSIS} is a radiative transfer simulation code provides simulated light curves for KNe model. This code generates light curves for NSBH as well as BNS model considering ejecta mass from dynamical and post-merger components of ejecta. A wide range of viewing angles from polar view ($\theta_{v}  = 0^\circ$) to equatorial view ($\theta_{v} = 90^\circ$) are considered by the code while generating the light curves. We used various possible combinations of ejecta mass from dynamical and post merger components: $\mathrm{M}(\rm{d}_{ej}) = [0.01, 0.03, 0.05]~\mathrm{M}_{\odot}$ and $\mathrm{M}(\rm{pm}_{ej}) = [0.03, 0.05, 0.09]~\mathrm{M}_{\odot}$ to compare our observations with simulated light curves. 

Figure \ref{fig:BH-NS-apr} depicts the simulated r$^{\prime}$ lightcurve for the NSBH models with various combinations of dynamical and post-merger ejecta masses. The rest-frame luminosity is converted into an apparent magnitude based on the distance to this source. The shaded band for each model denotes the 1-sigma range of distances, while the solid central line is calculated using the median distance of 353.2~Mpc \textit{for the region covered by GIT.} The corresponding absolute magnitudes are shown on the right side axis.
% with dynamical ejecta mass $\mathrm{M}(d_{ej}) = 0.01~\mathrm{M}_{\odot}$ (left panel), $0.03~\mathrm{M}_{\odot}$ (middle panel) and $0.05~\mathrm{M}_{\odot}$ (right panel) for post-merger ejecta masses $\mathrm{M}(\rm{pm}_{ej}) \in [0.03, 0.05, 0.09]~\mathrm{M}_{\odot}$. 
The red triangles denote the depth of GIT images. Since no counterpart was found, these indicate the upper limits to the brightness of a putative counterpart located in this part of the sky. Thus, if this is an NSBH event for which the true counterpart was in the region observed by GIT, we find that a scenario with $\mathrm{M(pm_{ej})} > 0.09~\mathrm{M}_{\odot}$ is very unlikely for polar viewing angles. For edge-on/equatorial view ($\theta_v = 90\degr$), the counterparts are expected to be fainter and evolve faster, and would be beyond the detection capabilities of GIT.

%Under the consideration that the kilonova associated to S190426c is a result of NSBH merger event and falls in localisation covered by GIT, we can rule out $\mathrm{M(d_{ej})} >~0.01~\mathrm{M}_{\odot}$ and $\mathrm{M(pm_{ej})}~>~ 0.09~\mathrm{M}_{\odot}$ for polar viewing angle, at a median distance of 353.2~Mpc calculated for localisation covered by GIT. Our upper limits are shallow to constrain the equatorial viewing angle models. 

\begin{figure*}
    \centering
        \includegraphics[width=0.9\linewidth]{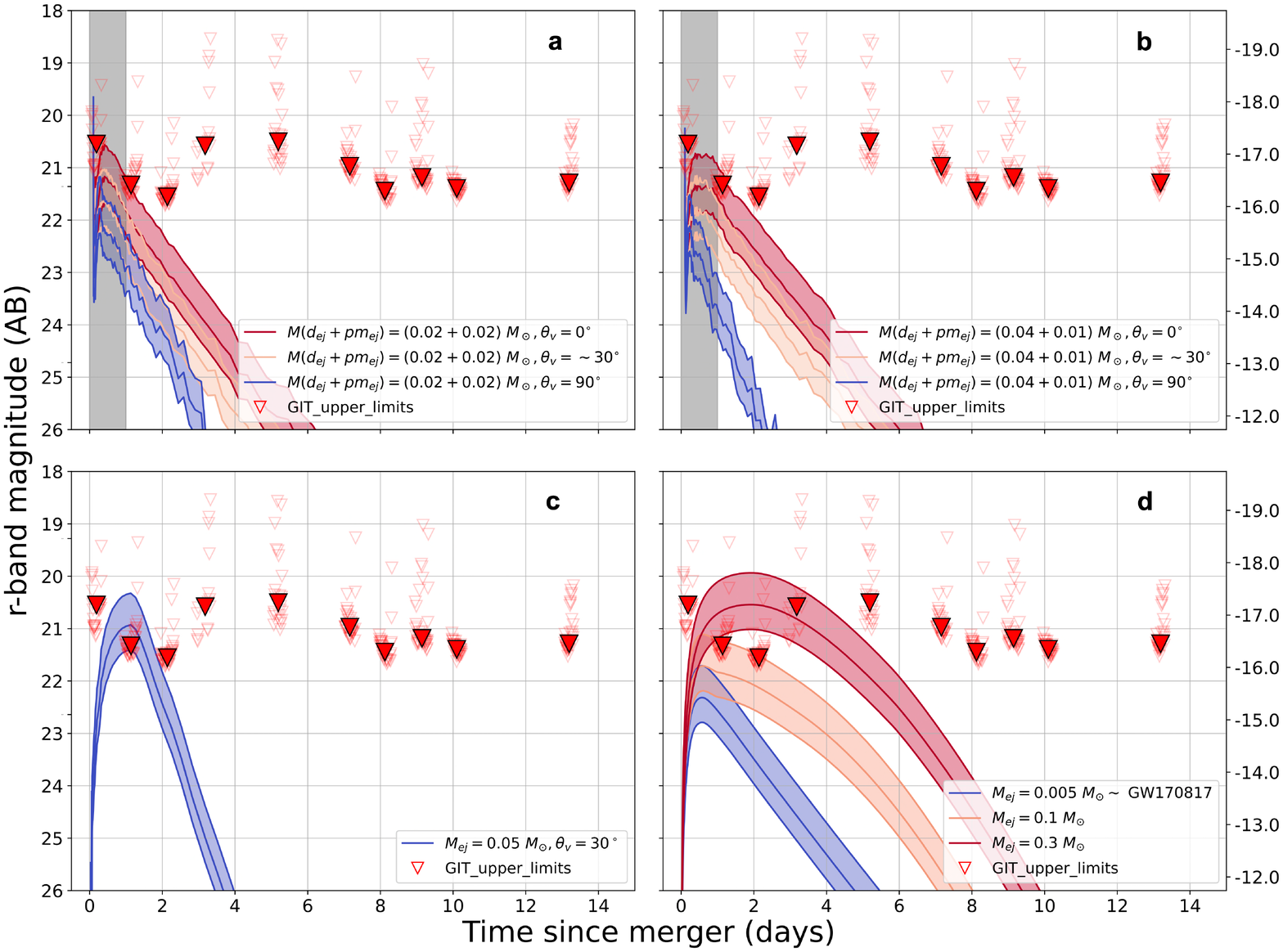} 
     \caption{Constraining source emission models using GIT data. The bands and magnitudes are as described in Figure~\ref{fig:KNe-models}. \textit{Panels a, b}: \citet{2020ApJ...889..171K} NSBH models, with blue denoting an equatorial viewing angle and redder colours indicating a polar view. The models are not very reliable for the first day, which is shown as a shaded gray region.
     \textit{Panel c}: Our data rules out the \citet{Banerjee_2020} blue KN model.
     \textit{Panel d}: We can rule out the high ejecta mass case for \citet{2020ApJ...891..152H} models (red curve), and constrain the distance to $\gtrsim 300$~Mpc for moderate ejecta mass (orange curve).}
     \label{fig:KNe-models_2}
\end{figure*}

We also considered NSBH KNe model by \citet{2020ApJ...889..171K} that explore the scenario of prompt collapse to form a black hole in compact object mergers using radiative transfer simulations (Figure \ref{fig:KNe-models_2}, upper panel). We select two combinations of dynamical and post-merger ejecta masses ($\mathrm{M}(\rm{d}_{ej}+\rm{pm}_{ej}) \leq {0.02 + 0.02}~\mathrm{M}_{\odot}$ and $\mathrm{M}(\mathrm{{d}_{ej}+{pm}_{ej}}) \leq {0.04 + 0.01}~\mathrm{M}_{\odot}$), seen at head-on and face-on inclinations seen from a range $\in$ [$0^{\circ}, 90^{\circ}$] of viewing angle are compared to our observations. These models predict a faster evolution of the counterpart, highlighting the importance of early observations. Note that the model is not very reliable during the first day post-merger (gray shaded region in Figure~\ref{fig:KNe-models_2}). Thus, the most important data is our upper limit at two days after the event - which disfavours an event with a polar viewing angle and high dynamical ejecta, located up to a distance of $\sim 300$~Mpc.

% However, we can rule out the polar angle view model for $\mathrm{M}(\rm{d}_{ej}+\rm{pm}_{ej}) \leq {0.04 + 0.01}~\mathrm{M}_{\odot}$ ejecta mass at nearest distance estimation for this event. 

\subsubsection{Binary Neutron Star models}
We now consider a series of BNS counterpart models and compare them to our upper limits.
Figure~\ref{fig:BNS-apr} shows the plotted simulated light curves for $\mathrm{M}(\rm{d}_{ej}) = [0.02, 0.01, 0.005]~\mathrm{M}_{\odot}$ for a very low and very high post-merger ejecta masses ($0.01 ~\mathrm{M}_{\odot}$ and $0.13 ~\mathrm{M}_{\odot}$ respectively), for polar and equatorial viewing angles.
%($\mathrm{M}(\rm{pm}_{ej}) = 0.01 ~\mathrm{M}_{\odot}$) and very high ($\mathrm{M}(\mathrm{pm}_{ej}) = 0.13~\mathrm{M}_{\odot}$) post merger ejecta mass in case of polar and equatorial viewing angles. 
GIT data can rule out the high post-merger ejecta mass cases for polar viewing angles (assuming the counterpart was located in the observed part of the sky). The low ejecta mass cases are ruled out if the source was located in the lower side of the allowed distances ($d \lesssim 300$~Mpc). We cannot strongly constrain the cases with low masses of post-merger ejecta.

%GIT $\rm r^\prime$ band upper limits put up strong constraints on high post merger ejecta mass for nearest distance estimate of 267.22~Mpc by LVC. The high post merger ejecta of ($\mathrm{M}(\rm{pm}_{ej}) = 0.13~\mathrm{M}_{\odot}$) is completely ruled out for polar viewing angle, while the moderate ejecta mass model ($\mathrm{M}(\rm{d}_{ej}+\rm{pm}_{ej}) \leq 0.01 + 0.13~\mathrm{M}_{\odot}$) is not plausible up to median distance of the event.
%
%For equatorial viewing angles, only the high dynamical ejecta mass model ($\mathrm{M}(\rm{d}_{ej}+\rm{pm}_{ej}) \leq 0.02 + 0.13~\mathrm{M}_{\odot}$) can be ruled out up to a distance of $\leq  300$~Mpc on the basis of upper limits around $\sim$ 2 and 3 days. GIT upper limits constraint ($\mathrm{M}(\rm{d}_{ej}+\rm{pm}_{ej}) \leq 0.03~\mathrm{M}_{\odot}$).

Next we consider the \citet{Banerjee_2020} blue KN model for BNS merger counterparts, which provides precise opacity calculations at early times.
%Another very popular early time blue kilonova model by \citet{Banerjee_2020} for BNS merger was compared with the early time upper limits obtained with GIT. This model provides precise opacity calculations at early times. 
We see that our observations one day after the merger can completely rule out the scenario with dynamical and post-merger ejecta masses of $0.02~\mathrm{M}_{\odot}$ and $0.05~\mathrm{M}_{\odot}$.

Lastly, we consider a KNe model by \citet{2020ApJ...891..152H}\footnote{These models are available at \url{https://github.com/hotokezaka/HeatingRate}}, very high ejecta mass model $\mathrm{M_{ej}} = 0.3~\mathrm{M}_{\odot}$ can be completely ruled out as shown in lower right panel of Figure~\ref{fig:KNe-models_2}. The $0.1~\mathrm{M}_{\odot}$ case can be constrained to require $d \gtrsim 300$~Mpc to be non-detectable by GIT. However, an event similar to GW170817  ($\mathrm{M_{ej}} = 0.005~\mathrm{M}_{\odot}$) at these distances would be undetectable by GIT.

In conclusion, GIT observations at early times can effectively rule out bright counterpart models, but are not deep enough to constrain the fainter models.

\section{Summary and future outlook}\label{sec:discuss}
% \outline{Observation stats: net coverage (area, prob, galaxies) --- image depth, seeing}
% \rough{Possible figure: histogram of image depths, histogram of seeing}
% \rough{Give \%\ completeness at lim-mag?}

%\subsection{Coverage}
% 
The GROWTH-India Telescope participated in a coordinated distributed campaign to search the localisation region of S190426c for electromagnetic counterparts. GIT observed the northern polar cap, with central and southern regions being observed by ZTF and DECam respectively. We covered 22.1~$\rm{deg}^2$ region of the sky, which had a 17.5\% probability of containing the counterpart. We created an alternate-night observing program and imaged the region for a total of ten days, with each point imaged multiple times so that we could trace the evolution of any candidate counterpart. We attained a typical 5-$\sigma$ limiting magnitude of 21.3 in our 10-minute exposures, and point sources had a medium FWHM of 3\arcsec. 

We imaged 8 NED and 332 glade galaxies, all of which were immediately visually searched for counterparts. We did not find any candidate counterparts in this search. We then developed a complete image subtraction and transient search pipeline, which was used to process the data and look for transients. Our tests show that we have a recovery rate of $\sim 94\%$ in such searches, though the actual ``raw'' efficiency was lower due to incomplete reference image coverage.

We discovered 23 transients, of which five were flagged as potential candidates while the rest were associated with stellar sources. The light curve evolution of showed that none of the candidates was consistent with being a counterpart of S190426c. We thus obtained upper limits on the peak flux, and hence the peak luminosity of a possible counterpart for the probability region covered by our observations. We compared our upper limits to various theoretical models for counterparts to BNS and NSBH mergers. We find that we can rule out models with high ejecta mass in all cases. Our upper limits are sensitive enough to rule out few more models in the closer part ($\lesssim 300$~Mpc) of the GW localisation volume. Our data are most useful in the first few days after the trigger, where the emission is expected to peak. Beyond $\sim5$ days after the GW event, most models predict significantly fainter emission which cannot be detected by GIT at large distances. Such continued follow-up with meter-class telescopes continues to be important for nearby events like GW170817.

%\subsection{Outlook for O4}
An increase in sensitivity of advanced LIGO detectors resulted in a significant increase in the detection of triggers with a non-zero probability of a neutron star as one of the merger objects: from a single event in O2 to fifteen triggers during O3~\citep{abbott2020gwtc2}. The larger sky areas and increased distances in the localisation volumes \citep{abbott2020gwtc2} necessitate a large number of images, each with increased depth. This poses a formidable challenge for smaller telescopes like GIT to cover significant portions of the sky regions. Increased sensitivity of LIGO and Virgo detectors as well as the participation of KAGRA in the subsequent observing runs will improve localisation for nearby events but will also add a large number of distant, poorly localised events \citep{2022ApJ...924...54P}. Telescope networks, and small telescopes in particular, need better strategies to deal with this scenario.

Follow-up of BNS and NSBH triggers remains a high priority for GIT, and we will invest significant time to triggers in the fourth GW observing run (O4) and beyond. The follow-up strategy --- tiling the localisation region, galaxy targeted search, photometric follow-up of external candidates --- will be evaluated on a case-by-case basis for each trigger. Our data processing and photometric pipelines are well-developed to enable rapid turn-around for targeted searches \citep{2022arXiv220613535K}. We are developing our image subtraction and transient search pipelines to increase our capabilities for blind searches for transients in our images. The improved pipelines will shorten the processing time, decrease the amount of human involvement needed, and more effectively discard spurious sources. Armed with these developments, we are confident that GIT will continue to play its role as a key resource for the electromagnetic follow-up of gravitational wave sources in the eastern hemisphere.

% In this section, we can summarise our work along with the type of mode of observation to go for in the future for small telescopes.

%Whether it is useful to go for the tiled follow-up using a small telescope? Any simulation that can be done for this shows what could be the essential criteria to trigger the small telescopes in what mode. This part needs some work to be done properly. 

%Given that the O4 run and ZTF \& Rubin observatory will observe, can this be a separate paper?

%\rough{What you are proposing is too detailed. Not for this paper.}
%We can compare our performance with other giant telescopes on a relative scale. How good/bad we have done.
%However, this might be seen as a counter to those other telescopes and should be avoided if there is any reason. I do not have any idea about it. Maybe you know it well. 
 
\section*{Acknowledgements}
The GROWTH India Telescope (GIT) is a 70-cm telescope with a 0.7-degree field of view, set up by the Indian Institute of Astrophysics and the Indian Institute of Technology Bombay with support from the Indo-US Science and Technology Forum (IUSSTF) and the Science and Engineering Research Board (SERB) of the Department of Science and Technology (DST), Government of India. It is located at the Indian Astronomical Observatory (Hanle), operated by the Indian Institute of Astrophysics (IIA). We acknowledge funding by the IITB alumni batch of 1994, which partially supports operations of the telescope. Telescope technical details are available at \url{https://sites.google.com/view/growthindia/}.

This research has made use of the NASA/IPAC Extragalactic Database (NED), which is funded by the National Aeronautics and Space Administration and operated by the California Institute of Technology.

This research has made use of data and/or services provided by the International Astronomical Union's Minor Planet Center. 

This research has made use of the VizieR catalogue access tool, CDS, Strasbourg, France (DOI : 10.26093/cds/vizier). The original description of the VizieR service was published in 2000, A\&AS 143, 23.
 
This research has made use of NASA's Astrophysics Data System.

Harsh Kumar thanks the LSSTC Data Science Fellowship Program, which is funded by LSSTC, NSF Cybertraining Grant \#1829740, the Brinson Foundation, and the Moore Foundation; his participation in the program has benefited this work.

MC acknowledges support from the National Science Foundation with grant numbers PHY-2010970 and OAC-2117997.

%%%%%%%%%%%%%%%%%%%%%%%%%%%%%%%%%%%%%%%%%%%%%%%%%%
\section*{Data Availability}
All data used in this article have been included in a tabular format within the article.



% \bibliographystyle{mnras}
% \bibliography{references} % if your bibtex file is called example.bib

%\begin{thebibliography}{99}
%\end{thebibliography}

%%%%%%%%%%%%%%%%%%%%%%%%%%%%%%%%%%%%%%%%%%%%%%%%%%

%%%%%%%%%%%%%%%%% APPENDICES %%%%%%%%%%%%%%%%%%%%%

% \appendix

% \section{Some extra material}

% NA
\bibliographystyle{mnras}
\bibliography{ref} 
% %%%%%%%%%%%%%%%%%%%%%%%%%%%%%%%%%%%%%%%%%%%%%%%%%%

% % Don't change these lines
% \bsp	% typesetting comment
\label{lastpage}
\end{document}